\documentclass[twocolumn,10pt,prl,aps]{revtex4-1}

\usepackage{siunitx}

\begin{document}

\title{Interlayer Configurations in Twisted Bilayers of Folded Graphene}

\author{J. C. Rode}
\author{C. Belke}
\author{H. Schmidt}
\author{R. J. Haug}

\affiliation{Institut f\"ur Festk\"orperphysik, Leibniz Universit\"at Hannover, 30167 Hannover}

\date{\today}

\begin{abstract}
The folding of monolayer graphene leads to new layered systems, termed twisted bilayer graphene (TBG), generally displaying a certain interlayer rotation away from crystallographic alignment. We here present an atomic force microscopy study on folded graphene, revealing unexpectedly large twist angle dependent modulations of $\sim\SI{3}{\angstrom}$ in interlayer distance. At the TBG surface, we find enhanced friction attributable to superlubricity in between incommensurate layers. At the bended edge, the radius of curvature scales with the folded length, congruent to earlier studies on carbon nanotubes.\\
\end{abstract}

\maketitle

Two dimensional crystals exhibit a unique variety of mechanical\cite{Novoselov2005,Xu2013,Annett2016,Lee2010,Lee2009,Chen2015}, electronic\cite{Xu2013,Geim2007,dosSantos2007} and optical\cite{Xu2013,Yin2016} properties. Even more complex systems are created by stacking of different materials\cite{Geim2013} or introduction of a rotational mismatch in fewlayer structures\cite{dosSantos2007}. Most prominent among the latter is twisted bilayer graphene, the electronic properties of which have been intensively studied in the past\cite{dosSantos2007,Brihuega2012,Yamagishi2012,Schmidt2014,Yin2016}. However details on TBG morphology remain largely unexplored: Reliable data on the interlayer distance between two graphene sheets for example, are in fact limited to the AB-stacking configuration. Predictions for simple AA-stacking remain unverified due to its metastable nature. Height measurements on twisted configurations focus on local corrugation, invoked by the periodic Moir\'e pattern between twisted layers\cite{Campanera2007,Brihuega2012} and are furthermore mostly conducted via scanning tunneling microscopy, thereby being highly dependent on density of states.  
To resolve these issues, we here present a comprehensive Atomic Force Microscope (AFM) study on the interlayer configuration in TBG produced by folding of monolayers. Interlayer distance shows a pronounced dependence on rotational mismatch and corresponding superlattice configurations\cite{Campanera2007,Mele2010,Uchida2014,vanWijk2015}; maximal observed layer separations exceed theoretical prediction significantly. Lateral force microscopy finds enhanced friction on TBG with respect to mono- and single-crystal fewlayer graphene, breaking the established decreasing trend over layer number\cite{Lee2010}. Unique to folded TBG, the radius of curvature at the bended edge is found to decrease with its length, which is relatable to earlier studies on carbon nanotubes\cite{Sinnott1999,Rochefort1999}.

As precursor to TBG, graphene monolayers are prepared by mechanical exfoliation of natural graphite onto silicon dioxide. By choice of this amorphous substrate, any anisotropic phenomena can safely be assumed to be intrinsic to TBG, which is in contrast to the popular choice of hexagonal boron nitride (hBN)\cite{Wang2016}. While the exfoliation procedure itself yields a small number of folded TBG to begin with, more folds are induced on demand by controlled cuts into monolayer flakes\cite{Annett2016}, using an AFM with a wear resistant tip operating in the \si{\micro N}-range (see schematic in Fig. 1(a)). 

\begin{figure}[hbtp]
\includegraphics{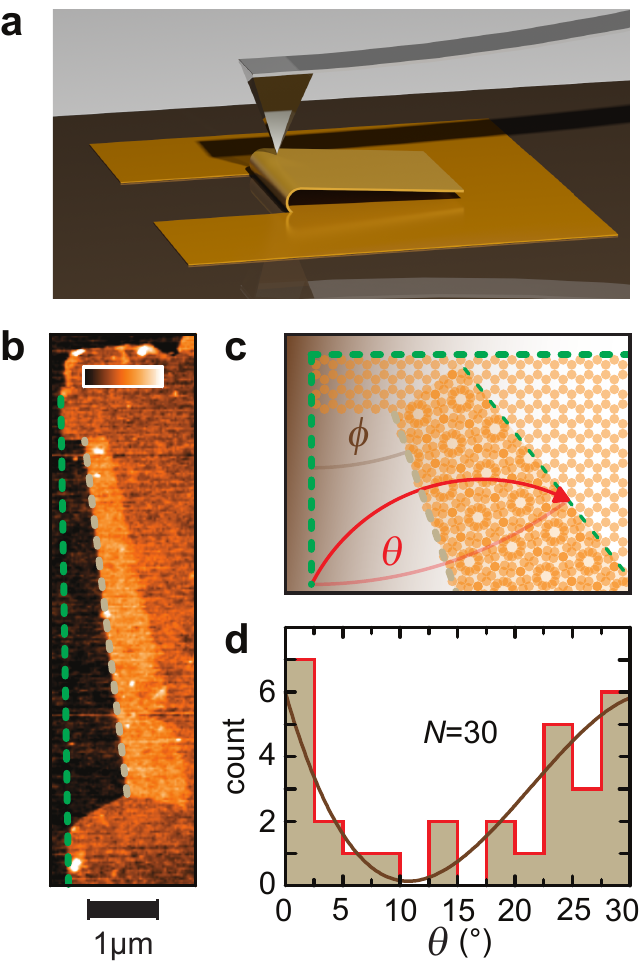}{\centering}
\caption{(a) Schematic of graphene (orange) on a substrate (brown), folded with an AFM-tip. (b) AFM topography of a folded graphene monolayer. Dashed green and beige lines trace straight sample edge and folded edge respectively, color scalebar spans \SI{2.25}{nm} in height. (c) Schematic of a folded graphene monolayer with crystallographic edges (dashed green), folded edge (dashed beige) and angles $\phi$ and $\theta$. The folded region displays a superlattice pattern. (d) Frequency of examined samples as function of twist angle in \SI{2.5}{\degree} increments. A polynomial fit of third order (dark brown line) serves as guide to the eye.}
\end{figure}

AFM topography of an example result is shown in Fig. 1b. Due to prevalently crystallographic edges in a graphene flake\cite{Neubeck2010}, the interlayer rotation angle $\theta$ can be conveniently deduced as twice the angle $\phi$ between folded edge and a straight sample edge as illustrated in Fig. 1(c). Figure 1(d) shows a histogram over thusly extracted angles projected into a range of \SI{0}{\degree} to \SI{30}{\degree}, which reveals concentrated occurrence at large and small twist angles with only few TBG of intermediate $\theta$. A very similar angular distribution has recently been observed in TBG grown by chemical vapor deposition\cite{Yin2016}. Studies on folded layers of HOPG\cite{Li2006} and thin carbon samples in liquid suspension\cite{Zhang2010} found a preference of angles around zero degrees only, which might be due to comparatively large edge to bulk ratio and corresponding impact of the folded edge.

Besides the twist angle $\theta$, interlayer distance $\Delta h$ is a defining parameter in TBG, being of obvious importance for band structure calculations and also interlayer capacitance. Most theoretical treatments assume an invariant value of $\Delta h_{AB}=\SI{3.35}{\angstrom}$, known from AB-stacked graphene and graphite\cite{dosSantos2007}. Experimentally a variety of different step heights have been found\cite{Geim2007,Yamagishi2012,Annett2016} but not yet systematically investigated. We here perform AFM topography measurements to examine $\Delta h$ in TBG of various known $\theta$. Height differences are extracted from histograms over topography information (see Figs. 2(a),(b)) and plotted over the interlayer twist angle in Fig. 2(c). Notably the single-crystal interlayer distance $\Delta h_{AB}\sim\SI{3.4}{\angstrom}$, as measured across a step between monolayer and AB-stacked bilayer graphene marks the lower limit of $\Delta h$ in TBG (see blue horizontal line). This is to be expected as Bernal- or AB-stacking is the energetic optimum of interlayer configurations\cite{Campanera2007, Uchida2014, Berashevich2011, Shibutaa2011}. Above this value however, $\Delta h$ evolves as function of $\theta$ over a range of $\sim\SI{3}{\angstrom}$, which will be discussed in the following. 

\begin{figure}[hbtp]
\includegraphics{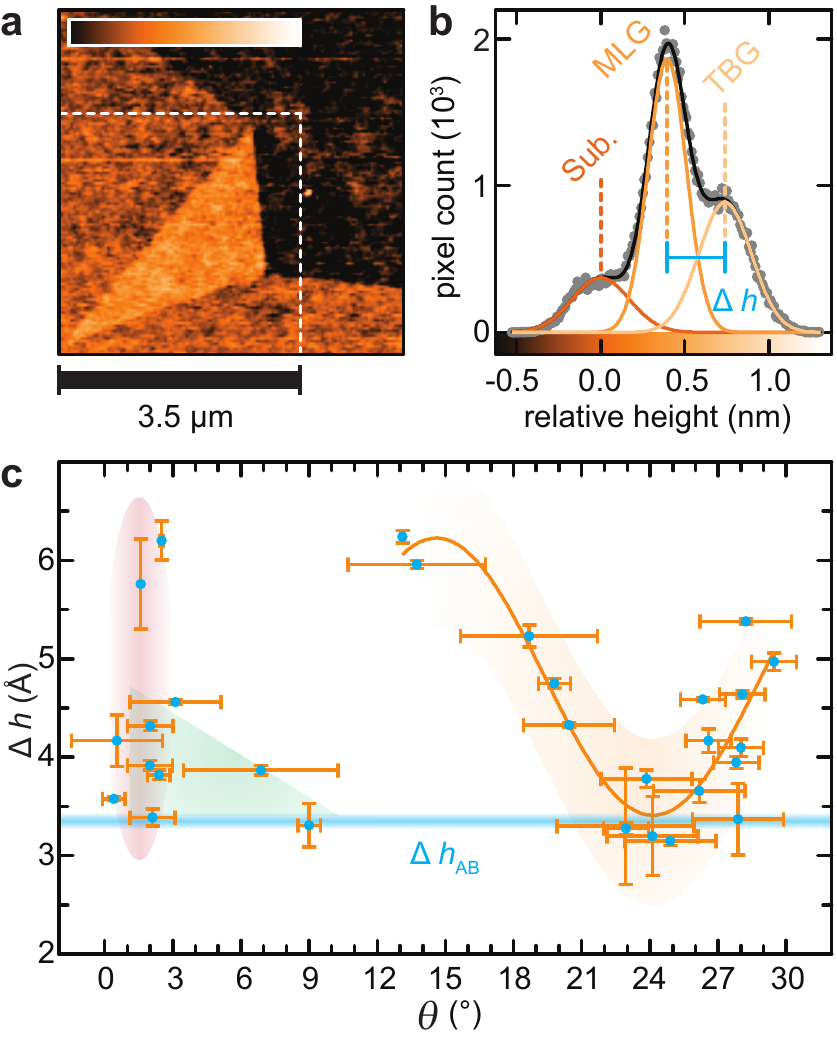}{\centering}
\caption{(a) AFM topography of a folded graphene monolayer. The white dotted square indicates area of analysis for panel (b), color scalebar spans \SI{2}{nm} in height. (b) Histogram of pixel frequency as a function of height (gray dots). Three contributions from substrate (Sub.), monolayer (MLG) and twisted bilayer graphene (TBG) can be distinguished and described by Gaussian distributions (colored lines) the sum of which (black line) is fit to the data. The interlayer distance $\Delta h$ as plotted in (c) is indicated in blue. (c) Blue dots: $\Delta h$ as function of interlayer twist $\theta$. Blue line: measured step height $\Delta h_{AB}$ between two layers of graphene in AB-stacking. Red area indicates scattering linked to MP corrugation $\Delta m$ at small $\theta$; green area indicates $\Delta m$ from ref.\cite{Brihuega2012}, offset by $\Delta h_{AB}$. A sinusodial fit (orange line) serves as guide to the eye for the oscillating behavior at larger $\theta$. Error bars in $\Delta h$ stem from fitting uncertainty, error bars in $\theta$ from measurement uncertainty.}
\end{figure}

To understand bulk interaction in TBG, formation of $\theta$-dependent Moir\'e pattern (MP) as well as commensurate superlattices structures have to be considered. In general\cite{Campanera2007,vanWijk2015}, two interposed honeycomb lattices of constant $a=\SI{246}{\pico m}$ lead to twist angle dependent MP of wavelength 
\begin{equation}
\label{eq:1}
\lambda(\theta)=a/(2\cdot\sin(\theta/2).
\end{equation} 
Within a corresponding Moir\'e unit cell, lattice registry alternates between areas similar to AB-stacking and congruent AA-stacking respectively (compare dark and light spots in Fig. 1(c)). In the energetically favored AB-like spots, $\Delta h$ will arrange close to $\Delta h_{AB}$ while AA-areas tend to buckle apart by an amount of $\Delta m$\cite{Berashevich2011,Brihuega2012,Uchida2014}, locally maintaining larger interlayer distance; a phenomenon known as Moir\'e pattern corrugation. 
At small $\theta$, these corrugations are most pronounced, furthermore undergoing qualitative changes in terms of their particular shape within the first \SI{1}{}-\SI{2}{\degree} of rotational mismatch\cite{vanWijk2015}. In the small-angle range (red area , Fig. 2(c)), the observed scattering of $\Delta h$ is therefore ascribed to both complexity of TBG morphology and measuring technique: As the typical radius of AFM tips employed in this work is $t_{r}\approx\SI{10}{nm}$, height measurements in close contact will be on the verge of averaging over a unit cell with $\lambda(\theta)\sim t_{r}$ for $\theta\sim\SI{1.5}{\degree}$ (compare eq. 1), thus yielding $\Delta h$ between $\Delta h_{AB}$ and $\Delta h_{AB}+\Delta m$ depending on shape and size of MP corrugation and sharpness of the AFM tip.

At intermediate angles, where $\lambda(\theta)<t_{r}$, deflection of the AFM scanning probe will mostly be dictated by corrugation \textit{maxima} only. In fact we here find good agreement with scanning tunneling measurements\cite{Brihuega2012} on MP modulation $\Delta m$, offset by $\Delta h_{AB}$, as indicated by the light green area in Fig. 2(c).  

At larger rotational mismatch of $\theta\gtrsim\SI{10}{\degree}$, MP wavelength and corrugation become significantly smaller\cite{Uchida2014,Brihuega2012}, so interlayer distance can be treated as uniform. Furthermore, discrete commensurate angles with strictly periodic superlattice\cite{Campanera2007,Mele2010,vanWijk2015} are considerably less dense among the generally semi-periodic Moir\'e structures at larger $\theta$, which may have significant impact on TBG electronic structure\cite{Mele2010,Koren2016,Chari2016}. Height measurements in the according angular range reveal oscillating behavior over $\sim\SI{3}{\angstrom}$ with a pronounced dip down to $\sim\Delta h_{AB}$ at $\theta\sim\SI{24}{\degree}$ (orange area in fig. 2c). Theoretical predictions for the planar $\Delta h$ are scarce, calculations for interaction energy $E_{int}$ contradictory: A virtually $\theta$-independent $E_{int}$ was found, based on Lennard-Jones interaction between lattice atoms\cite{Shibutaa2011}. DFT calculations predict a monotonous decrease of $E_{int}$ over $\theta$ on the order of $\sim\SI{1}{meV}$/atom\cite{Uchida2014,Campanera2007}. While none of the former references predict more than $\sim\SI{0.3}{\angstrom}$ in variation of $\Delta h$, a larger span of up to $\SI{1}{\angstrom}$ is found via sophisticated quantum chemistry methods\cite{Berashevich2011}. Although the precision of this value may be limited by the finite size of the simulated flake\cite{Berashevich2011}, it still cannot account for the span of $\sim\SI{3}{\angstrom}$, covered by our data. These discrepancies encourage further work on the understanding of interlayer configuration in TBG. In this, it should be noted that the observed dip in $\Delta h$ appears close to a predicted local energy minimum\cite{Berashevich2011} and the important commensurate angle $\theta\sim\SI{21.8}{\degree}$, which has very recently been associated with resistance minima in rotatable inter-graphitic junctions\cite{Koren2016,Chari2016}.                                       

Additional information about the manner of interlayer coupling may be deduced from mechanical behavior of the TBG surface, as measured via Lateral Force Microscopy (LFM). Beside the obvious impact of surface morphology, frictional energy dissipation in LFM is believed to be mainly influenced by out-of-plane elasticity\cite{Lee2010,Choi2011}. Decreasing friction from MLG to bulk graphite has been observed for increasing number of layers and is ascribed to decreasing pliancy in thicker samples\cite{Lee2010}.

\begin{figure}[hbtp]
\includegraphics{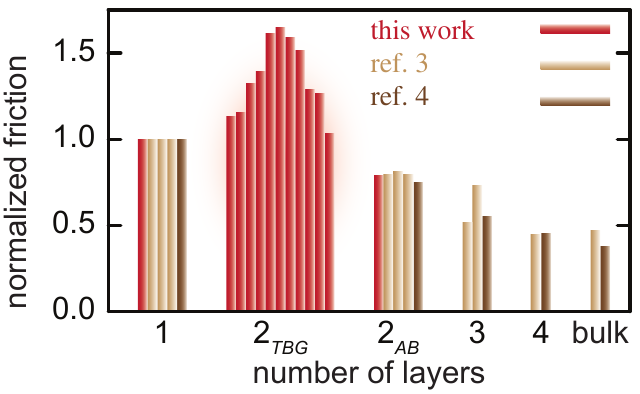}{\centering}
\caption{Normalized friction on monolayer, twisted bilayer and single-crystal multilayer graphene. Beige and brown bars represent data from refs.\cite{Lee2010,Lee2009} according to legend. Red data are step measurements, normalized for comparison as described in the methods section.}
\end{figure}
 
Figure 3 shows frictional data on our TBG, normalized to the friction of a graphene monolayer. In the same figure a measurement on an AB-stacked bilayer and data of single-crystal multilayers from other references\cite{Lee2010,Lee2009} are plotted for comparison. The observable variation among TBG values is likely due to frictional anisotropy caused by hexagonal superlattice MP and strain-induced ripples\cite{Choi2011}, which could not be accounted for in our unidirectional step measurements. Interestingly though, friction on all of the examined TBG is clearly larger than on Bernal-stacked bilayers and even monolayers, breaking the decreasing trend over layer number. Following ref.\cite{Lee2010} this points towards enhanced ductility of graphene in twisted stacking-configuration, which has three possible explanations: Firstly, additional out-of-plane elasticity may locally be caused by MP corrugation which will furthermore provide increased surface area. Secondly, weaker interlayer binding with respect to Bernal-configuration may allow for local detachment of the twisted top layer and dynamic deformation in response to a scraping AFM tip. Lastly, the TBG as a whole is likely to be more ductile than its Bernal-stacked counterpart: Examining folded fewlayer graphene, bending rigidity was found to strongly depend on interlayer sheer interaction\cite{Chen2015}. As graphitic interfaces become superlubric upon rotational misalignment\cite{Dienwiebel2004}, TBG top and bottom layer will shear freely, thereby increasing the bilayer´s ductility with respect to crystallographically aligned AB-stacking. Our findings thus suggest a counter-intuitive causality between interlayer superlubricity and enhanced LFM friction at the TBG surface.   

Finally, the bended edge connecting top and bottom layer in our TBG is unique to the folding approach in preparation. It is predicted to give rise to snake states\cite{Rainis2011} and has been linked to transport features independent of in-plane magnetic field\cite{Schmidt2014}. The shape of the folded edge depends on bending stiffness and adhesion between TBG planes\cite{Chen2015,Rainis2011,Cranford2009} and is described by a bending radius $r$ (see schematic in Fig. 4(b)). It plainly shows in AFM topography as a bump of height $\Delta b$ (see Fig. 4(a),(c)) approximately related to $r$ and interlayer distance $\Delta h$ via 
\begin{equation}
\label{eq:2}
r=(\Delta h+\Delta b)/2.
\end{equation} 
We find bump heights between \SI{0}{\angstrom} and \SI{8.5}{\angstrom} (Fig. 4(c)) and calculate corresponding bending radii between \SI{1.7}{\angstrom} and \SI{6.3}{\angstrom} (Fig. 4(d)). The radius of curvature shows no apparent dependence on interlayer twist, which indicates isotropic bending stiffness in congruence with theoretical prediction\cite{Cranford2009}. Interestingly though, $r$ systematically increases with length $\ell$ of the folded edge. A decreasing slope (see gray line in Fig. 4(d) as guide to the eye) suggests saturation against an estimated radius $r_{sat}$ between $6$ and $\SI{7}{\angstrom}$. Note that these values compare very well to theoretical prediction\cite{Rainis2011} of $r_{theo}\approx\SI{7}{\angstrom}$, working in disregard of the third dimension parallel to the fold. Interestingly, the behavior at shorter bended edges can be linked to calculations on carbon nanotubes, which find increasing stability for larger tube lengths\cite{Sinnott1999,Rochefort1999}. This would render a short bended area more prone to deformation or even rupture upon shifting in the bulk; furthermore a tendency to larger tube diameters for bigger systems is predicted due to the energetic struggle between strain of curvature and number of edge atoms\cite{Sinnott1999}. As a folded edge may be seen as half a carbon nanotube, our findings are qualitatively in line with the discussed predictions and bring together two important subjects in chemistry and physics.

\begin{figure}[hbtp]
\includegraphics{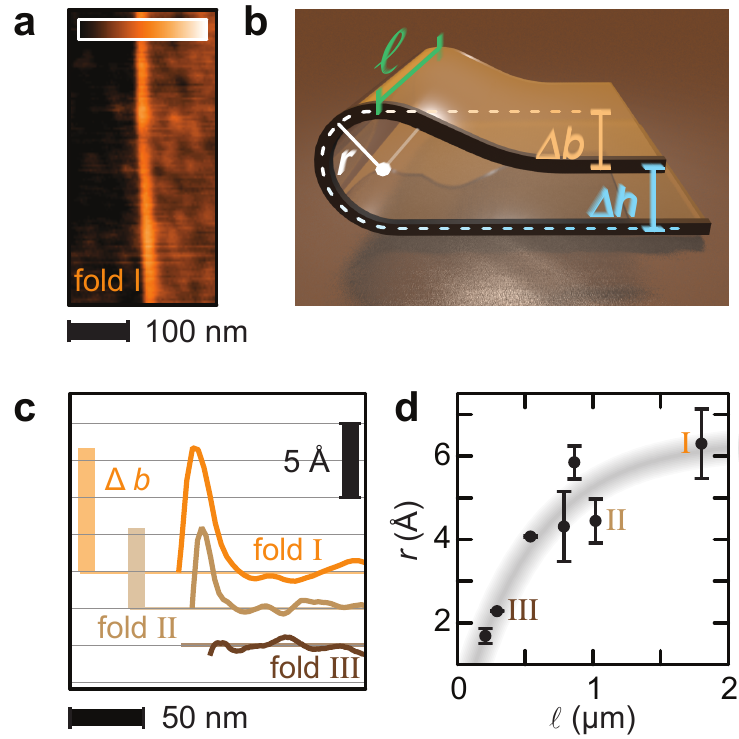}{\centering}
\caption{(a) AFM topography along a folded edge corresponding to the orange cross section in (c). Color scalebar spans \SI{3}{nm}. (b) Schematic of a folded graphene flake, illustrating the relations expressed in equation (2). The green line points in the third dimension, spanning the length $\ell$ of the folded edge. (c) Averaged height profiles across three folded edges with different bump heights. Traces are offset by \SI{2.5}{\angstrom} for better visibility. (d) Bending radius vs. length of folded edge. Folds in (b) are identified by roman numerals, an exponential fit (gray line) serves as guide to the eye.}
\end{figure}

In summary, twisted bilayers prepared by folding of monolayer graphene are found to preferentially arrange around twist angles of \SI{0}{\degree} and \SI{30}{\degree}. In interlayer distance, an unexpectedly large variation over $\sim\SI{3}{\angstrom}$ is observed and compared to a number of different coupling models in terms of angular dependence. Enhanced friction on the twisted bilayer surface with respect to single-crystalline stacking is attributed to increased pliancy and superlubricity between twisted layers. Specific to folded samples, the radius of curvature at the interconnection between top and bottom layer is examined and found to increase with length of the folded edge, which is in conjunction with predictions on carbon nanotubes.   

\section{Acknowledgment}

The authors acknowledge financial support from the DFG within the priority program SPP 1459 and the School for Contacts in Nanosystems.

J. C. Rode acknowledges support from the Hannover School for Nanotechnology.

The authors thank Peter Behrens for fruitful discussion.

\section{Appendix}
As the majority of edges in a graphene flake is terminated by armchair or zigzag edges\cite{Neubeck2010} and the folded edge acts as an axis of reflection between top and bottom layer, $\theta$ may be calculated as twice the angle $\phi$ between folded edge and crystallographic direction. Due to graphene´s sixfold symmetry, the obtained angle may be projected into a range of $\theta\in[\SI{0}{\degree},\SI{30}{\degree}]$ ($\theta$ and $\SI{60}{\degree}-\theta$ being identical but for a possible translational shift depending on axis of rotation). The accuracy of this geometric method usually lies around $\pm\SI{1.5}{\degree}$ depending on the original edge measurements.

AFM height measurements were performed in contact mode, taking special care to minimize the effects of friction-induced mechanical crosstalk, which might otherwise corrupt step measurements across heterogeneous materials\cite{Warmack1994}. Height differences are determined via a histogram of pixel frequency over recorded topography information, which is fit by the sum of \textit{n} Gaussians, \textit{n} being the number of expected height levels. $\Delta h$ is extracted as the difference between mean values of TBG and MLG distributions, uncertainty is set to the sum of corresponding fitting uncertainties.

Similar to step heights $\Delta h$ in topography, we extract TBG friction $\Delta V$ with respect to an adjoining monolayer. For comparison with measurements on single crystal fewlayers\cite{Lee2010,Lee2009}, data are normalized via the step $\Delta V_{AB}$ from  monolayer to AB-stacked bilayer and offset by 1.0.


\newpage
\end{document}